\documentclass[amsmath,amssymb,twocolumn,amsfonts]{revtex4-2}
\usepackage{graphicx}
\usepackage{epsfig}
\usepackage{bm}
\usepackage{bbm,mathbbol}
\usepackage{braket,bigints}
\usepackage{amsthm}
\usepackage{physics}
\usepackage{siunitx}
\usepackage{enumitem}
\usepackage{hyperref}
\usepackage{stmaryrd,mathtools}
\usepackage{csquotes}
\hypersetup{colorlinks=true,linkcolor=blue,citecolor=blue,urlcolor=blue}
\newcommand{\EE}{\mathbb{E}}

\newcommand{\br}{\mathbf{r}}
\newcommand{\bx}{\mathbf{x}}
\newcommand{\by}{\mathbf{y}}

\newcommand{\bJ}{\mathbf{J}}

\def \beq {\begin{equation}}
\def \eeq {\end{equation}}
\def \tr {\rm Tr}

\begin{document}
\title{Metabolic quantum limit to the information capacity of magnetoencephalography}
\author{E. Gkoudinakis$^{1,2}$}
\author{S. Li$^3$}
\author{I. K. Kominis$^{1,2}$}
\email{ikominis@uoc.gr}
\affiliation{$^1$School of Science, Zhejiang University of Science and Technology, Hangzhou 310023, China}
\affiliation{$^2$Department of Physics, University of Crete, Heraklion 70013, Greece}
\affiliation{$^3$School of Automation and Electrical Engineering, Zhejiang University of Science and Technology, Hangzhou 310023, China}

\begin{abstract}
Magnetoencephalography measures the magnetic fields generated by neural currents using quantum sensors such as superconducting quantum interference devices and atomic magnetometers. Here we combine the energy resolution limit of magnetic sensing with the metabolic power available to neural currents to derive a technology-independent bound on the information capacity of MEG. The bound factorizes into geometry, metabolism, and Planck's constant, and gives an estimated maximum information rate of 2.2~Mbit/s for representative human-brain parameters. Further, we show that the externally measurable magnetic field has a finite angular bandwidth, with high multipole components being geometrically attenuated and falling below the quantum-limited noise floor. This yields an information-limited spatial scale of order $1~{\rm cm}$ and renders the accessible measurement space effectively finite-dimensional. The energy resolution limit therefore defines an information-theoretic Nyquist scale for magnetoencephalography, beyond which denser spatial sampling provides redundant measurements rather than additional recoverable information. Since the energy resolution limit also makes the noise variance grow linearly with measurement bandwidth, temporal and spatial bandwidths compete, producing a fundamental spatio-temporal trade-off. These results show how quantum-limited measurements constrain the observable complexity and information content of noninvasive brain imaging, providing a quantitative link between fundamental physics and neuroscience.
\end{abstract}
\maketitle 
\section{Introduction}
Magnetoencephalography \cite{Cohen1972,Sato1991,Ilmoniemi1993,Knowlton1997,Salmelin1996,Hari1999,Vrba2001,Gratta2001,Nolte2003,Wheless2004,Fokas2004,Hillebrand2005,Goldenholz2009,Stam2010,Hari2012,Baillet2017} presents a direct window into the human brain, capturing neural activity through the magnetic fields generated by ionic currents in the cortex. The development of magnetoencephalography (MEG) paralleled that of superconducting quantum interference devices (SQUIDs) \cite{ClarkeRMP,Vrba2002,Fagaly2006,Yang2009,Oisjoen2012,Faley2013}, which long provided the sensitivity required to detect femtotesla-level brain signals. More recently, atomic magnetometers achieved comparable sensitivities without cryogenics \cite{Xia2006,Johnson2013,Kim2014,Alem2014,Colombo2016,Boto2017,Sheng2017,Borna2017,Boto2018,Guo2018,Tierney2019,Iivanainen2019,Barry2019,He2019,Zhang2019,Labyt2019,Guo2020,Borna2020,Vivekananda2020,Rea2022,Feys2022}. As quantum sensing technologies become increasingly integrated into neuroscience and medicine, it becomes essential to identify the fundamental physical principles that ultimately bound the information capacity of MEG.

Recent work has established a technology-independent energy resolution limit (ERL) for magnetic sensing \cite{MitchellERL,KominisERL}, demonstrating that the product of magnetic-field variance, sensor volume, and measurement time is bounded from below by Planck's constant. The ERL provides a subtle and unifying framework for analyzing diverse magnetic sensing experiments. For example, it was recently applied to elucidate the quantum limits underlying biological magnetoreception \cite{Kominis2025}.

Here we derive a fundamental bound on the information capacity of MEG that is independent of sensor technology and measurement configuration. By linking the ERL with the metabolic energy expended by transmembrane ion pumps, we obtain a bound that factorizes into three irreducible elements: geometry (encoded in the continuous lead-field structure), metabolism (the energetic cost of sustaining neural currents), and quantum physics ($\hbar$). Beyond bounding information, we show that the same principles impose a fundamental limit on MEG spatial resolution, moreover intertwining spatial and temporal bandwidth. 

The term \enquote{metabolic} is not merely semantic. It points to a general paradigm: biochemical energy  $\rightarrow$ physical observable  $\rightarrow$ quantum-limited measurement. MEG thus emerges as a platform for probing how the laws of quantum measurement constrain information flow in biological systems. This perspective can inspire further connections of modern quantum technology to neuroscience \cite{Wakai1996,Frederick2006,Ioannides2006,Parkkonen2009,Cichy2017,Gross2019,Cichy2015} and medical diagnosis \cite{Rose1987,Knowlton2004,Makela2006,Stufflebeam2009,Mohamed2013,Wilson2016}.

The paper is organized as follows. In Sec.~II we formulate MEG in the continuum limit by introducing the lead-field covariance operator and its spectrum, which provides a sensor-independent description of the spatial correlations of the measured magnetic field. In Sec.~III we combine this continuum information measure with the energy resolution limit and the metabolic power available to neural currents, obtaining a bound on the MEG information capacity and evaluating it for representative human-brain parameters. In Sec.~IV we show that the same framework implies a finite spatial bandwidth and a fundamental spatio-temporal trade-off, leading to an information-limited spatial resolution. Finally, in Sec.~V we discuss the robustness of the bound, its interpretation in terms of finite accessible modes and sampling, and its implications for MEG instrumentation and inverse reconstruction.
\section{Lead Field Covariance Operator in the Continuum}
The information conveyed by MEG is so far quantified by considering a discrete sensor array in specific configurations \cite{Kemppainen,Iivanainen2017,Schneiderman2014,Skidchenko2023}. We treat the {\it continuum case}, where the entire volume outside the head is probed. Consider a spherical volume $V$ wherein exist current dipoles described by the density $\mathbf{J}(\bx)$, generating a magnetic field in the space $\Omega$ extending beyond the volume $V$, and being separated by a distance $d$ from the boundary of $V$. 
At position $\br$ within $\Omega$ the magnetic field components are (with $\alpha,\beta=x,y,z$) $B_\alpha(\br) = \int_V L_{\alpha\beta}(\br,\bx)\,J_\beta(\bx)\,d\bx$, with $\vb J$ representing the primary currents. The lead field $L_{\alpha\beta}(\br,\bx)$ is given analytically by the Sarvas model (see Appendix A), applicable to arbitrary radial conductivity $\sigma=\sigma(r)$ in the source region \cite{Geselowitz1967,Geselowitz1970,Sarvas1987,Mosher1999}.

The lead-field operator ${\cal L}$ and its adjoint ${\cal L}^\ast$ are: 
\begin{align}
{\cal L}[\mathbf{J}]_\alpha(\br) &= \int_V  L_{\alpha\beta}(\br,\bx)\,J_\beta(\bx)\,d\bx,\label{lfop}\\
{\cal L}^\ast [\mathbf{f}]_\beta(\bx)&=\int_\Omega L_{\alpha\beta}(\br,\bx)\,f_\alpha(\br)\,d\br.
\end{align}
Then the lead-field covariance operator is 
\beq
{\cal K}_\Omega={\cal L}{\cal L}^\ast,
\label{KOmega}
\eeq
with kernel $K_{\alpha\beta}(\br,\br') = \int_V L_{\alpha\gamma}(\br,\bx)L_{\beta\gamma}(\br',\bx)\,d\bx$. All spatial correlations of the magnetic field are contained in ${\cal K}_\Omega$. Because $L_{\alpha\beta}^2\!\sim\!|\br|^{-6}$ in the far field and $\Omega$ is separated from~$V$ by a finite distance $d$, the integral $\int_\Omega\!\int_V L_{\alpha\beta}^2\,d\bx\,d\br$ converges. Hence ${\cal K}_\Omega$ is a compact, positive trace-class operator with eigenvalues $\kappa_\ell>0$ (having units ${\rm T^2m^4/A^2}$), and $\sum_\ell\kappa_\ell<\infty$. 

We now consider random current dipoles with delta-correlated current densities: 
\beq
\EE[J_\alpha(\bx)J_\beta(\by)]=\mathbb{V}[J]\,\delta_{\alpha\beta}\,\delta(\bx-\by),
\label{J}
\eeq
where $\mathbb{V}[J]$ is the variance of the current-dipole density (having units ${\rm A^2/m}$), and is the same for all three Cartesian components. Even though the underlying current sources are random, the magnetic field exhibits spatial correlations due to the structure of the lead fields. Moreover, there is additive sensor noise contributing to the measured signal $\tilde{\mathbf{B}}(\br)=\mathbf{B}(\br)+\mathbf{b}(\br)$. The noise field has covariance 
\beq
\EE[b_\alpha(\br)\,b_\beta(\br')] =\mathbb{V}[b]\delta_{\alpha\beta}\,\delta(\br-\br'),
\label{eq:noise}
\eeq
where $\mathbb{V}[b]$ reflects the magnetic field estimate variance, taken equal for all three Cartesian components ($\mathbb{V}[b]$ has units ${\rm T^2m^3}$). Thus, $\tilde{B}_\alpha(\br) = \int_V L_{\alpha\beta}(\br,\bx)\,J_\beta(\bx)\,d\bx + b_\alpha(\br)$ describes a linear map from random currents to random fields. Assuming $\bJ$ and $\mathbf{b}$ are Gaussian, $\tilde{\mathbf{B}}$ is also Gaussian. We now evaluate the information conveyed by $\tilde{\mathbf B}$. The source and noise covariance operators are
\begin{align}
\EE[\bJ\otimes\bJ]&=\mathbb{V}[J]\mathcal{I}_V,\label{EJJ}\\
\EE[\mathbf{b}\otimes\mathbf{b}]&=\mathbb{V}[b]\mathcal{I}_\Omega, \label{Ebb}
\end{align}
where $\mathcal{I}_V$ and $\mathcal{I}_\Omega$ are the identity operator in the Hilbert space of square-integrable vector fields defined in $V$ and $\Omega$, respectively (see Appendix A). 

Let $\{\mathbf{u}_\ell\}$ be an orthonormal eigenbasis of ${\cal K}_\Omega$, ${\cal K}_\Omega[\mathbf{u}_\ell](\br)=\kappa_\ell \mathbf{u}_\ell(\br)$, $\langle \mathbf{u}_\mu,\mathbf{u}_\ell\rangle=\delta_{\mu\ell}$, where the inner product is $\langle \mathbf f,\mathbf g\rangle= \int_\Omega \mathbf f(\mathbf r)\!\cdot\!\mathbf g(\mathbf r)\,d\br$. The projection of the measured field onto $\{\mathbf{u}_\ell\}$ is
\beq
\tilde{\beta}_\ell=\langle \tilde{\mathbf B},\mathbf{u}_\ell\rangle
= \langle {\cal L}[\mathbf J],\mathbf{u}_\ell\rangle + \langle \mathbf b,\mathbf{u}_\ell\rangle\label{beta_m}
\eeq
Using independence and zero mean, we have $\EE[\tilde{\beta}_\ell]=0$ and $\EE[\tilde{\beta}_\ell^2]=\EE\!\left[\langle {\cal L}[\mathbf J],\mathbf{u}_\ell\rangle^2\right]+\EE\!\left[\langle \mathbf b,\mathbf{u}_\ell\rangle^2\right]$. Using \eqref{KOmega} and \eqref{EJJ} we get $\EE\!\left[\langle {\cal L}[\mathbf J],\mathbf{u}_\ell\rangle^2\right]=\mathbb V[J]\;\langle \mathbf{u}_\ell,{\cal L}{\cal L}^\ast \mathbf{u}_\ell\rangle=\mathbb V[J]\;\kappa_\ell$, and from \eqref{Ebb}, $\EE\!\left[\langle \mathbf b,\mathbf{u}_\ell\rangle^2\right]
= \mathbb V[b]\langle \mathbf{u}_\ell, \mathbf{u}_\ell\rangle=\mathbb V[b]$. Therefore,
\beq
\EE[\tilde{\beta}_\ell^2]=\mathbb V[J]\kappa_\ell+\mathbb V[b]\label{beta_ell}
\eeq
Interpreting the two terms in \eqref{beta_ell} as \enquote{signal} and \enquote{noise}, Shannon's formula provides the total information obtained from a spatially continuous measurement over $\Omega$,
\beq
I=\frac12\sum_{\ell}\log_2\!\left(1+\frac{\mathbb V[J]}{\mathbb V[b]}\,\kappa_\ell\right)
\label{eq:I}
\eeq
The eigenvalues $\kappa_\ell$ quantify how efficiently random currents in $V$ excite {\it independent} magnetic-field modes $\mathbf{u}_\ell(\mathbf r)$ in $\Omega$. The total information $I$ is finite because $\sum_\ell\kappa_\ell<\infty$, a consequence of the compactness of ${\cal K}_\Omega$. The continuous operator $\mathcal{K}_\Omega=\mathcal{L}\mathcal{L}^\ast$ thus provides a self-contained, sensor-independent description of magnetic-field correlations in $\Omega$, determined solely by the lead field geometry. Its spectrum sets an upper bound on the information accessible to any finite array, while avoiding convergence issues in discrete limits related to sensor modeling, noise normalization, or weighting.
\section{Bound on the MEG Information Capacity Enforced by the Energy Resolution Limit of Magnetic Sensing and Brain's Metabolic Power}
The utility of the ERL will now become apparent. For a magnetic field estimate of variance $(\delta B)^2$ over volume $v$ with bandwidth $W$, the energy resolution per unit bandwidth is $\epsilon=(\delta B)^2v/2\mu_0W$, and the ERL states that $\epsilon\gtrsim\hbar$. We assume access to all three Cartesian components of the field with independent, identically distributed noise as in \eqref{eq:noise}, so that this bound applies to each component separately. In the continuum limit $v\to 0$, the product $(\delta B)^2v$ remains finite \cite{MitchellERL}, and we identify it with $\mathbb{V}[b]$. Thus $\mathbb{V}[b]\gtrsim 2\mu_0\hbar W$, hence the sought-after MEG capacity (information in bits/s) reads
\beq
I_W\leq{W\over 2}\sum_{\ell}\log_2\!\left(1+\frac{\mathbb V[J]}{2\mu_0\hbar W}\kappa_\ell\right)\label{bound1}
\eeq

In the last step of our derivation we connect $\mathbb{V}[J]$ with the metabolic power, $P_{\rm mb}$, driving the current dipoles. Like the standard approach \cite{Plonsey1982}, we consider only intra-dendrite axial currents during excitatory postsynaptic potentials (EPSP) as the dominant contribution to MEG signals. Such currents consist of sodium and calcium ions flowing inward and potassium ions flowing outward, resulting in a net transmembrane current $I_0$. We approximate the affected dendritic segment with a semi-infinite cylindrical cable of cross section $A_d$ and intracellular (axial) resistivity $\rho$ (since MEG signals are dominated by intra-dendritic axial currents, the relevant dissipation occurs within neuronal processes rather than the extracellular bulk tissue \cite{Wu}). For a localized injection of $I_0$ at $x=0$, the solution of the cable equation is an axial current $I(x)=I_0 e^{-x/\lambda}$, where $\lambda$ is the electrotonic length. This axial current translates into a local current density $J(x)=I(x)/A_d$. Because this current decays exponentially, points separated by more than $\lambda$ are effectively uncorrelated. The correlation volume is $V_c=\int e^{-2x/\lambda}A_ddx=A_d\lambda/2$. 

Now, the ohmic dissipation in the volume $V$ is $\int \rho J^2(\br)d\br$. By identifying this dissipation with $P_{\rm mb}$, we get for the spatial average of $J^2$, $\langle J^2\rangle_V=(1/V)\int J^2d\br=P_{\rm mb}/\rho V$. To connect this coarse-grained quantity to the delta-normalized correlation used in Eq. \eqref{J} we multiply by the correlation volume $V_c$, obtaining $\mathbb{V}[J]=P_{\rm mb}A_d\lambda/2\rho V$. In physical terms, $P_{\mathrm{mb}}/(\rho V)$ gives the height of the current density's spatial correlation function at zero separation, while $V_c$ represents the width of that correlation. Replacing the finite-range correlation by a delta function should preserve its total integral, which leads to the expression above. Substituting into \eqref{bound1} yields 
\beq
I_W\leq{W\over 2}\sum_{\ell}\log_2\!\left(1+{{P_{\rm mb}A_d\lambda}\over {2\rho V}}{\kappa_\ell\over {2\mu_0\hbar W}}\right)\label{bound2}
\eeq
This is the first main result of this work. As stated in the introduction, it is cast in terms of three factors: a geometric ($\kappa_\ell$), reflecting the geometric coupling of current dipoles into magnetic fields, a metabolic ($P_{\rm mb}$), reflecting the energy driving the current dipoles, and $\hbar$. The information $I_W$ grows with $W$. If the current dipole source is band-limited to $B_J$, increasing the sampling rate $W$ beyond $2B_J$ does not increase $I_W$. Therefore, the MEG information capacity is given by \eqref{bound2} for $W=2B_J$.

To estimate $P_{\rm mb}$ from the metabolic energy budget \cite{Attwell2001,Lennie2003,Raichle2010,Howarth2010,Howarth2012,Harris2012,Clarke2013,Engl2015,Engl2017,Yu2017,Jamadar2025}, we first evaluate the total transmembrane current $I_0$. With $E_{\mathrm{Na}}\approx +60~\mathrm{mV}$ being the sodium equilibrium potential, and a sodium driving force of about $120~\mathrm{mV}$, the membrane potential is $V\approx -60~\mathrm{mV}$ (since $E_{\mathrm{Na}}-V\approx 120~\mathrm{mV}$). For potassium it is $E_{\mathrm{K}}\approx -90~\mathrm{mV}$, so its driving force is about $|V-E_{\mathrm{K}}|\approx 30~\mathrm{mV}$. Assuming equal conductances for Na$^+$ and K$^+$, the currents scale with their driving forces and have opposite directions, giving $I_{\mathrm{Na}}\approx -4 I_{\mathrm{K}}$. Using $I_{\mathrm{Na}}^{\mathrm{NMDA}}\approx 0.64~\mathrm{pA}$ (inward, with 10\% carried by Ca$^{2+}$) and $I_{\mathrm{Na}}^{\mathrm{AMPA}}\approx 32~\mathrm{pA}$ (inward) \cite{Attwell2001} yields $I_{\mathrm{Na}}\approx 32.6~\mathrm{pA}$ inward and $I_{\mathrm{K}}\approx -8.2~\mathrm{pA}$ outward, so the net current is $I_0=I_{\mathrm{Na}}^{\mathrm{NMDA}}+I_{\mathrm{Na}}^{\mathrm{AMPA}}+I_{\mathrm{K}}\approx 24.5~\mathrm{pA}$ inward. The intra-dendritic resistance for a semi-infinite dendritic cable is $R_{\mathrm{in}}=\rho\lambda/A_d$. With $A_d=\pi(0.5~\mu\mathrm{m})^2$, $\lambda=0.3~\mathrm{mm}$, and $\rho=1~\Omega\mathrm{m}$  \cite{Wu}, this gives $R_{\mathrm{in}}\approx 380~{\rm M\Omega}$. The ohmic power dissipated during an EPSP is then $P_{\rm epsp}=\tfrac{1}{2}R_{\mathrm{in}} I_0^2\approx 115~\mathrm{fW}$. The corresponding energy, $E_{\rm epsp}\approx 0.115~\mathrm{fJ}$, dissipated over the 1 ms duration of the EPSP \cite{Attwell2001}, is a small fraction of the free energy resulting from the hydrolysis of $\sim1.4\times10^{5}$ ATP molecules \cite{Clarke2013}. Summed over $\sim 1.1\times 10^{10}$ excitatory cortical neurons, each with $\sim 10^4$ successful synaptic events per second \cite{Yu2017}, the sought-after power is $P_{\mathrm{mb}}= 12~\mathrm{mW}$. 

For a source region (cortex) of radius $a$ and a separation $d$ between the cortical surface and the measurement region, i.e. an observation region $\Omega$ consisting of points $\mathbf r$ such that $|\mathbf r|\ge a+d$, the eigenvalues of ${\cal K}_\Omega$, derived in Appendix A, are $\kappa_\ell=\mu_0^2 a^2 \frac{\ell}{(2\ell+1)^2(2\ell+3)} (\frac{a}{a+d})^{2\ell+1}$, with $\ell=1,2,...$, and they are $(2\ell+1)$-degenerate. Using the above with $W=1000~{\rm Hz}$, $a=8~{\rm cm}$, and $d=1.3~{\rm cm}$, we find $I_W\lesssim 2.2~\mathrm{Mbit/s}$. 

We emphasize that introducing metabolic parameters allows our result to inherit the remarkable stability of the brain's global energy budget, which remains constant within 5\%  regardless of transient cognitive demands \cite{Jamadar2025, Raichle2010}. Clearly, this numerical estimate does not claim to be any more precise than the numerical values used to derive it. Also, realistic (non-spherical) head geometries \cite{Iivanainen2017} might change this number. Nevertheless, it is evident that there is quite some room for acquiring more information with MEG, since state-of-the-art systems report capacities $\sim$400 bits per sample \cite{Schneiderman2014}, which for the same sampling rate translate to 0.4 Mbit/s. 

These results are visualized in Fig. 1, where we plot the eigenvalues $\kappa_\ell$ (Fig. 1a), and the behavior of $I_W$ (Fig. 1b) as a function of the number of eigenvalues included in the sum in the RHS of \eqref{bound2}. For the latter plot we use an ERL which ranges from $\hbar$ (fundamental limit) to $10^4\hbar$, so that we cover a broad sensitivity range.
\begin{figure}
\begin{center}
\includegraphics[width=8.5 cm]{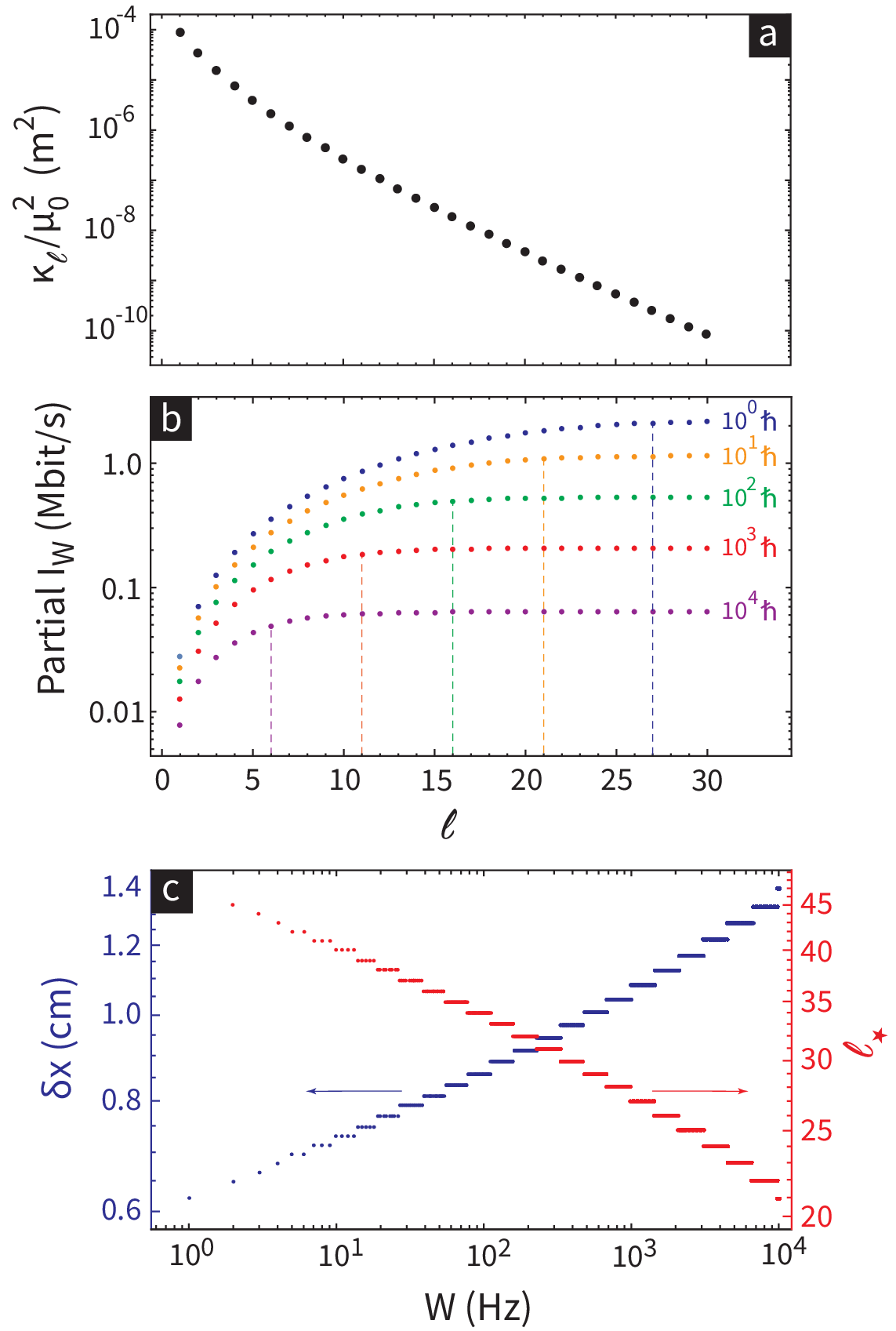}
\caption{(a) The first 30 eigenvalues of ${\cal K}_\Omega$ normalized by $\mu_0^2$, for $a=8~{\rm cm}$ and $d=1.3~{\rm cm}$. (b) Partial information obtained from \eqref{bound2} by letting the sum over $\ell$ extend from 1 to $m=1,2,...,30$, and generalizing the ERL to be in the range from $\hbar$ (the fundamental limit) up to $10^4\hbar$. It is seen that the more sensitive the sensor, the more spatial modes of ${\cal K}_\Omega$ are required to saturate $I_W$. The dotted lines indicate the respective values of $\ell_\star$, quantifying the spatial resolution for various values of the energy resolution. (c) Spatial resolution (left $y$-axis) and mode number cutoff $\ell_\star$ (right $y$-axis) as a function of sampling rate $W$.}
\label{head}
\end{center}
\end{figure}
\section{Bound on MEG Spatio-Temporal Resolution}
Importantly, the above formalism also yields a quantitative limit on spatial resolution. There exists a maximal mode number $\ell_\star$, defined as the smallest $\ell$ for which the signal-to-noise ratio in \eqref{beta_ell} satisfies ${\rm SNR}_\ell \le 1$, or equivalently, using \eqref{bound1}, $\kappa_\ell \le 2\mu_0 \hbar W/\mathbb V[J]$. Modes with $\ell>\ell_\star$ contribute negligibly to the total information. The corresponding angular resolution is $\delta\theta \approx \pi/\ell_\star$.

Since outside the source region the magnetic field is harmonic, $\vb B=-\nabla\Phi$ with $\Delta\Phi=0$, its exterior multipole expansion is uniquely determined by the spherical-harmonic coefficients on any enclosing sphere. Radial continuation to larger radii merely rescales each angular component of $\Phi$ by a factor $r^{-(\ell+1)}$ and does not introduce additional angular degrees of freedom. Hence all independent angular information is already encoded on the innermost accessible sphere of radius $a+d$, and the associated linear spatial resolution at that surface is $\delta x \approx \pi (a+d)/\ell_\star$. 

Because the singular system of the lead-field operator preserves angular momentum, the same cutoff $\ell_\star$ also limits the angular complexity of current dipole patterns that contribute to the measurable magnetic field. Referring the resolution back to the cortical surface at radius $a$ yields the source-scale resolution $\delta x_{\mathrm{src}} \approx \pi a/\ell_\star$. That is, even with continuous spatial coverage and quantum-limited sensors, only current patterns up to a finite angular complexity, set by $\ell_\star$, leave a detectable imprint above the noise floor. Finer spatial modes may exist physically, but they are information-theoretically inaccessible. For our chosen parameters, and sensors operating at the ERL, we obtain $\ell_\star = 27$, giving $\delta x\approx\delta x_{\mathrm{src}}\approx 1~\mathrm{cm}$. Smaller values of $\ell_\star$ are obtained for sensors operating above the ERL (see dashed lines in Fig.~1b).  We stress that $\delta x$ quantifies the maximum spatial variability of current sources that leaves a detectable imprint in the measured magnetic field, and should not be confused with current dipole localization accuracy (see Appendix B).

It is also worth emphasizing that higher spherical-harmonic components decay rapidly with distance from the source region. A mode of angular index $\ell$ generated at radius $r$ contributes to the external field at sensor radius $R$ with an amplitude scaling approximately as $(r/R)^{\ell}$. Thus, high-$\ell$ angular structure generated deep within the brain is attenuated before reaching the sensors. The external magnetic field therefore carries only a finite angular bandwidth of information, even in the absence of detector noise. Combined with the ERL-imposed noise, this geometric attenuation leads to the spatial bandwidth cutoff $\ell_\star$ derived above. In other words, the cutoff $\ell_\star$ defines an effective information-limited rank of the forward operator. Although modes with $\ell>\ell_\star$ are not strictly silent, they fall below the ERL-limited detectability threshold and cannot be inferred from the data, thereby constraining the inverse problem to a finite data-supported subspace.

The ERL can further sharpen the understanding of spatial resolution limits. Resolution-matrix analyses \cite{Wens2023} typically assume fixed sensor noise, under which increasing the number of sensors can improve spatial discrimination. 
In contrast, the ERL fixes a minimum magnetic-field noise spectral density in the continuum limit. Since this implies a nonzero noise variance over any finite bandwidth, the accessible angular content remains finite and is set by the cutoff degree $\ell_\star$.

Interestingly, the cutoff depends explicitly on the measurement bandwidth $W$, as shown in Fig.~1c, through its defining condition $\kappa_{\ell_\star} \sim 2\mu_0\hbar W/\mathbb V[J]$. Since $\kappa_\ell$ decays exponentially with $\ell$ via the geometric factor 
$(a/(a+d))^{2\ell+1}$, increasing $W$ raises the effective noise floor and shifts the intersection with the spectrum $\{\kappa_\ell\}$ toward smaller $\ell$. Consequently, $\ell_\star$ decreases and $\delta x$ increases (spatial resolution worsens). Under the ERL scaling $\mathbb V[b]\propto W$, temporal bandwidth and spatial bandwidth therefore compete. Using the large-$\ell$ asymptotics of $\kappa_\ell$, one finds $\ell_\star \propto \ln(\text{const}/W)$, so that $\delta x \propto 1/\ln(\text{const}/W)$: the trade-off is logarithmic but fundamental.

This trade-off must be interpreted relative to the intrinsic source bandwidth $B_J$. If $W<2B_J$, increasing $W$ raises the total information rate $I_W$ but reduces the spatial bandwidth by lowering $\ell_\star$. Once $W\ge 2B_J$, further increasing $W$ adds no signal content while increasing the ERL-imposed noise variance, thereby degrading spatial resolution without improving $I_W$. The optimal choice is therefore $W=2B_J$, which maximizes information while preserving the largest spatial bandwidth allowed by the ERL.
\section{Discussion}
Overall, the ERL imposes a finite cutoff degree $\ell_\star$ that effectively restricts the rank of the forward operator. Consequently, while the magnetic field exists as a spatial continuum, the accessible field is band-limited and finite-dimensional in the angular momentum representation. Spherical sampling theory implies that it can be reconstructed from a finite number of independent spatial samples. Hence, to extract statistically independent and physically meaningful information, sensor spacing on the order of $\delta x\approx 1~{\rm cm}$ is sufficient. In this sense, the ERL sets an information-theoretic Nyquist scale for MEG measurements. We emphasize that we start from the continuum to obtain a sensor-independent description of the measurement. By working with the continuous lead-field operator and its covariance, the finite spatial bandwidth and associated granularity emerge as consequences of geometry and quantum-limited noise, rather than being imposed by an assumed discretization. 

For the representative value $\ell_\star=27$, the number of independent angular modes is $N_{\rm modes}=\sum_{\ell=1}^{\ell_\star}(2\ell+1)=\ell_\star(\ell_\star+2)\approx \ell_\star^2\approx 700$. This also sets the scale for the required sensor count: to capture all resolvable spatial degrees of freedom without aliasing, the number of independent sensors must be on the order of $N_{\rm modes}$ \cite{Elahi2018}. Moreover, sensor volume and spacing are not independent: both must be matched to the intrinsic scale $\delta x$, since a finite sensor volume acts as a spatial low-pass filter. If its linear dimensions approach or exceed $\delta x$, spatial averaging suppresses the highest accessible modes and effectively lowers the cutoff $\ell_\star$, reducing the number of recoverable modes.

Importantly, while the quantitative value of the bound depends on modeling assumptions, its qualitative structure is robust. The existence of a finite spatial bandwidth and an effective cutoff $\ell_\star$ arises from the interplay of geometric attenuation and the energy resolution limit, and therefore persists under more realistic source models. Our analysis assumes spatially and temporally uncorrelated current sources in order to derive a maximum achievable information rate. Incorporating realistic correlations, such as spatial network structure or $1/f$ temporal dynamics, would reduce the effective number of independent degrees of freedom and thus tighten the bound, without altering its fundamental structure.

The present framework yields additional insights providing guidance for instrumentation. The dependence of the cutoff $\ell_\star$ on bandwidth establishes a fundamental trade-off between temporal and spatial resolution: increasing the measurement bandwidth raises the quantum-limited noise floor and reduces the accessible spatial complexity. Thus, to maximize the accessible information, the measurement bandwidth should match the intrinsic source bandwidth, $W \approx 2B_J$. Once the spatial and temporal sampling criteria are met, further gains cannot be achieved by increasing sensor density, but only by reducing measurement noise. In this sense, the ERL shifts the focus of MEG design from increasing sensor density to optimizing sensitivity within a fundamentally finite-dimensional measurement space.

In conclusion, this work shows how quantum-limited measurements fundamentally constrain the observable complexity of brain activity, providing a bridge between neuroscience, the physics of information, and quantum sensing technology, while also pointing toward future human–machine interface paradigms \cite{Zhu2020} in which high-fidelity neural signals can be harnessed for non-invasive communication and control.
\appendix
\section{Trace and Eigenvalues of ${\mathcal K}_\Omega$ in the Sarvas Model}
In the quasistatic regime, for a spherically symmetric conductor of conductivity $\sigma = \sigma(\rho)$ inside and $\sigma = 0$ outside, the magnetic field outside of the conductor is given by the Sarvas formula \cite{Sarvas1987}.

\paragraph{The model}

The forward mapping from primary currents in $V=\{|\vb x|\le a\}$ to the magnetic field in $\Omega=\{|\vb r|\geq a+d\}$ is obtained by superposition of dipolar Sarvas fields, ${\cal L}[\mathbf{J}]_\alpha(\br) = \int_V  \mathsf L_{\alpha\beta}(\br,\bx)\,J_\beta(\bx)\,d\bx$, with
\beq
\mathsf L_{\alpha\beta}(\br,\bx) = 
\frac{\mu_0}{4\pi F^2} \left[ (\partial_\alpha F) \epsilon_{\beta\gamma\delta} r_\gamma x_\delta - F \epsilon_{\alpha\beta\gamma} x_\gamma \right]
\label{eq:sarvas_kernel}
\eeq
where $F = s (rs + r^2 - (\vb x \cdot \vb r))$ and $\nabla_{\vb r} F = (\frac{s^2}{r} + \frac{\mathbf{s} \cdot \vb r}{s} + 2s + 2r)\vb r- (s + 2r + \frac{\mathbf{s} \cdot \vb r}{s}) \vb x $ for $\mathbf{s} = \vb r - \vb x$ and $s=|\mathbf{s}|$.

The trace of the field covariance operator is given by the \emph{Hilbert--Schmidt norm}
\begin{equation}
\tr\{\mathcal {K}_\Omega\}
   = \int_{\Omega}\!\int_{V}
     \|\mathsf L(\vb r,\vb x)\|_{F}^{2}\,d\br\,d\vb x,
\end{equation}
where $\|\mathsf L\|_{F}$ denotes the Frobenius norm
$\|\mathsf L\|_{F}^{2} = \sum_{\alpha,\beta} |\mathsf L_{\alpha\beta}|^{2}$.
The result is, with $R=a+d$,
\begin{align}
\tr\{\mathcal {K}_\Omega\} &= \frac{\mu_0^2}{4} \left[
\frac{(3R^2-a^2)}{2}\ln\left(\frac{R+a}{R-a}\right)-3aR
\right] \nonumber\\
&= \mu_0^2 a^2\sum_{\ell=0}^\infty \frac{\ell}{(2\ell+1)(2\ell+3)} (a/R)^{2\ell+1}
\label{eq:newtrace}
\end{align}

\paragraph{Rotational symmetry and eigenfunctions}

The Sarvas kernel is rotationally covariant:
for any $\mathsf R\in{\rm SO}(3)$, $\mathsf L(\mathsf R\vb r,\mathsf R\vb x)=\mathsf R\,\mathsf L(\vb r,\vb x)\,\mathsf R^{T}$
since $F(\vb r,\vb x)$ is rotationally invariant and $\nabla_{\vb r}F(\vb r,\vb x)$ transforms covariantly under rotations. So, $\mathcal K_\Omega$ commutes with the ${\rm SO}(3)$ action. Therefore $\mathcal K_\Omega$ decomposes into irreducible angular-momentum subspaces $\mathcal X_\ell$ (the spaces of the poloidal modes),
and by Schur's lemma
\begin{equation}
\mathcal K_\Omega|_{\mathcal X_\ell}=\kappa_\ell I,
\qquad \text{with multiplicity }(2\ell+1).
\label{eq:schur}
\end{equation}
This property ensures that $\mathcal{L}$ maps the angular momentum subspace of order $\ell$ in the source domain strictly to the corresponding subspace of order $\ell$ in the field domain.

Outside the conductor, Maxwell's equations give $\nabla\times\vb B=0$ and $\nabla\cdot\vb B=0$ in $\Omega$, so the natural, regular at infinity, eigenfields are the exterior harmonic (poloidal) modes $ \vb B_{\ell m}(\vb r)=-\nabla (r^{-(\ell+1)} Y_{\ell m}(\hat r) )$ for $\ell \ge 1$. 

In a spherically symmetric conductor, the external field is generated only by the tangential (toroidal) component of the primary current. Radially oriented components belong to the null space and do not contribute to the external magnetic field. Thus, the source space is spanned by the toroidal harmonics $\vb J_{\ell m}(\vb x) = \vb x \times \nabla(\rho'^\ell Y_{\ell m}(\hat x))$. Therefore, the Sarvas kernel admits the following spectral decomposition
\begin{equation}
   \mathsf L(\vb r, \vb x) = \sum_{\ell,m}{ c_{\ell m}\vb B_{\ell m}(\vb r) \otimes \vb J_{\ell m}(\vb x)}
\end{equation}

\paragraph{Eigenvalues}

For $\vb B_{\ell m}$, $|\vb B_{\ell m}|\sim \rho^{-(\ell+2)}$, hence
\begin{equation}
\|\vb B_{\ell m}\|_{\Omega}^{2}
\sim
\int_R^\infty \rho^2 \,|\vb B_{\ell m}(\rho)|^2\,d\rho
\propto
\frac{R^{-(2\ell+1)}}{2\ell+1}.
\label{eq:normOmega_scaling}
\end{equation}
Moreover, 
\begin{equation}
    \mathcal L^\ast [\vb B_{\ell m}](\vb x) = \sum_{\ell',m'} c_{\ell' m'}{\vb J}_{\ell' m'}(\vb x)\int_\Omega \vb B_{\ell' m'}(\vb r) \cdot \vb B_{\ell m}(\vb r) d\vb r
\end{equation}
which, due to spherical-harmonic orthogonality, reduces to
\begin{equation}
    \mathcal L^\ast [\vb B_{\ell m}](\vb x) =  c_{\ell m}||\vb B_{\ell m}||_\Omega^2 \vb J_{\ell m}(\vb x)
\end{equation}
with
\begin{equation}
    ||\mathcal L^\ast [\vb B_{\ell m}](\vb x)||_V^2 = |c_{\ell m}|^2 ||\vb B_{\ell m}||_\Omega^4 ||\vb J_{\ell m}(\vb x)||_V^2 \sim \frac{a^{2\ell +3}}{R^{4\ell+2}}.
\end{equation}
Finally, using
\begin{equation}
\kappa_\ell=\frac{\|\mathcal L^\ast [\vb B_{\ell m}]\|_V^2}{\|\vb B_{\ell m}\|_\Omega^2} \sim  a^2 (\frac{a}{R})^{2\ell+1}.
\label{eq:kappa_ratio}
\end{equation}
and 
\begin{align}
\tr\{\mathcal{K}_\Omega\} &= \mu_0^2 a^2 \sum_{\ell=0}^\infty \frac{\ell}{(2\ell+1)(2\ell+3)}({a\over R})^{2\ell+1}\nonumber\\
&= \sum_{\ell=0}^\infty{(2\ell+1)\kappa_\ell},
\end{align}
we identify the eigenvalues of the field covariance operator
\begin{equation}
\kappa_\ell= \mu_0^2 a^2\frac{\ell}{(2\ell+1)^2(2\ell+3)} (\frac{a}{R})^{2\ell+1},\ell \ge 1
\label{eq:kappa_ell}
\end{equation}\newline\newline
\noindent
\section{Spatial Resolution versus Source Localization Accuracy}
The spatial resolution $\delta x$ quantifies the spatial bandwidth of the magnetic-field measurement space. The eigenfunctions of $\mathcal K_\Omega$ are magnetic-field patterns in $\Omega$, and $\ell_\star$ marks the highest angular degree whose signal-to-noise ratio exceeds unity. In spherical geometry, the singular system of $\mathcal L$ establishes a correspondence between magnetic-field modes and source spherical harmonics of the same degree $\ell$. 
The cutoff $\ell_\star$ therefore limits the maximal angular complexity of current dipole distributions contributing to the accessible information.

This spatial bandwidth should not be confused with dipole localization accuracy. Localization refers to estimating the parameters of a specific low-dimensional source model, such as a single dipole, and can remain precise even when high-$\ell$ modes are suppressed. 
In the spherical model, the exterior multipole expansion of a dipole located at radius $r<a$ contains coefficients scaling as $(r/R)^\ell$. A small displacement $r\to r+\delta r$ modifies these coefficients as $\Delta c_\ell \propto \ell\,\frac{\delta r}{r}\left(\frac{r}{R}\right)^\ell$, so the exponential geometric suppression $(r/R)^\ell$ continues to dominate the scaling. Thus a small displacement primarily modifies the low-order harmonics, while higher-order contributions remain exponentially suppressed.

To quantify localization precision within the same continuum measurement model used in the main text, we evaluate the Fisher information through the noise-weighted $L^2(\Omega)$ norm of the field derivative. For a dipole of moment $m$ at distance $s$ from the sensors, the magnetic field scales as $B\sim \mu_0 m/s^3$, and using the exact angular structure of the dipole field one finds $\|\partial_s \mathbf B\|_\Omega^2= 9\mu_0^2 m^2/10\pi s^5$. The Fisher information for the radial position is therefore $I_F(s)=\|\partial_s \mathbf B\|_\Omega^2/\mathbb V[b]=9\mu_0^2 m^2/10\pi\mathbb V[b]s^5$. The Cram\'er-Rao bound yields
\[
\delta s \gtrsim {1\over \sqrt{I_F(s)}}=\sqrt{\frac{10\pi}{9}\,\frac{\mathbb V[b]\; s^5}{\mu_0^2 m^2}}.
\]
Imposing the energy resolution limit, $\mathbb V[b]\ge 2\mu_0\hbar W$, gives the quantum bound
\[
\delta s \gtrsim 
\sqrt{\frac{20\pi}{9}\,\frac{\hbar W\, s^5}{\mu_0\, m^2}}.
\]
This estimate uses the standard dipolar far-field scaling, which is consistent with the asymptotic behavior of the Sarvas solution. For representative values $s\approx 0.09\,{\rm m}$, $m\sim10^{-8}\,{\rm A\,m}$, and $W\sim 1\,{\rm kHz}$, this bound gives $\delta s\gtrsim 10^{-7}\,{\rm m}$, far below mm or cm scales pertinent to the spatial bandwidth. In practice, modeling errors and classical noise dominate, leading to millimeter-level localization, while the spatial bandwidth limit $\delta x$ may be of order centimeters. Analogous estimates for angular parameters likewise yield uncertainties much smaller than $\pi/\ell_\star$.

\end{document}